# BEAM PHYSICS OF INTEGRABLE OPTICS TEST ACCELERATOR AT FERMILAB

A. Valishev, S. Nagaitsev, Fermilab, Batavia, IL 60510, U.S.A.
D. Shatilov, BINP SB RAS, Novosibirsk, 630090, Russia
V. Danilov, ORNL, Oak Ridge, TN 37831, U.S.A.

*Abstract*

Fermilab's Integrable Optics Test Accelerator (IOTA) is an electron storage ring designed for testing advanced accelerator physics concepts, including implementation of nonlinear integrable beam optics and experiments on optical stochastic cooling. The machine is currently under construction at the Advanced Superconducting Test Accelerator facility. In this report we present the goals and the current status of the project, and describe the details of machine design. In particular, we concentrate on numerical simulations setting the requirements on the design and supporting the choice of machine parameters.

## IOTA GOALS AND DESIGN

The IOTA ring was designed for the proof-of-principle experiment of the nonlinear integrable optics idea [1] at the ASTA facility [2]. The initial version of the ring design described in [3] was comprised of four periodic cells and had full 8-fold mirror symmetry. It was later identified that it is desirable to accommodate more experiments, such as the nonlinear focusing with electron beam lens [4] and optical stochastic cooling. These options demanded incorporation of a 5 m-long straight section. The experimental hall dimensions allow to accommodate the stretched ring (Fig. 1) keeping the four 2 m nonlinear magnet insertions and the e- beam energy of 150 MeV. The straight section opposite to the long experimental insertion will be used for injection, RF cavity and instrumentation. The ring lattice is comprised of 50 conventional water-cooled quadrupole and 8 dipole magnets. The beam pipe aperture is 50 mm.

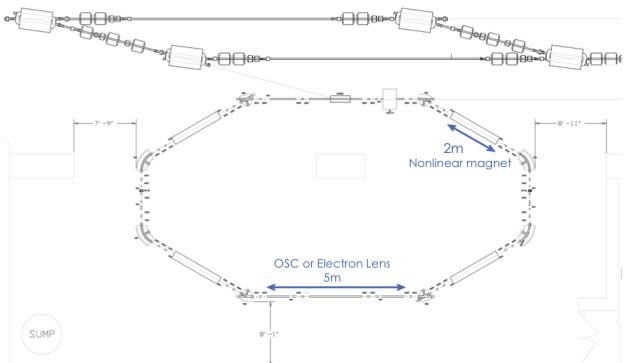

Figure 1: Layout of the IOTA ring.

The goal of experiments at IOTA is to demonstrate the possibility to implement nonlinear integrable system in a realistic accelerator design. The project will concentrate on the academic aspect of single-particle motion stability in the nonlinear integrable system, leaving the studies of collective effects and attainment of high beam current to future research [5]. We intend to achieve the amplitude-dependent nonlinear tune shift exceeding 0.25 without degradation of dynamic aperture.

Understanding of the limitations resulting from imperfections of the practical realization of the integrable optics is essential for the success of the project. The numerical tracking simulations discussed in this report were used to determine the design constraints.

## SIMULATION TECHNIQUE

The main goal of our simulations was to investigate how various misalignments and imperfections affect the integrability and the achievable tune spread in IOTA. We found the Frequency Map Analysis (FMA) [6] to be useful for this task, as it allows clear identification of important resonances, good and bad areas in the phase space, and the footprint shape. Since the technique uses single particle tracking, emittance of the stored beam does not matter, which agrees well with the assumed experimental approach of mapping the phase space with a pencil beam. In order to have a reasonable range of normalized betatron amplitudes, we set the emittance to $10^{-6}$ m. In this case the pole of nonlinear magnet is located at $A_x=10\,\sigma_x$ and the working area in the plane of normalized amplitudes $A_x$, $A_y$ will be $5 \times 15$.

First, we assume the machine lattice between nonlinear magnets to be linear and fully consistent with the requirement of an ideal axially symmetrical lens. In the current design of nonlinear magnets, the nonlinear potential will be approximated by a number of thin elements of constant cross-section. The nominal design calls for 20 slices for each magnet, and we used simulations to support this choice. FMA plots for the nominal case (20 slices) are presented in Fig. 2, where the betatron tune jitter along the particle trajectory is shown as a function of the initial amplitudes by colour in logarithmic scale: blue ($10^{-7}$) corresponds to stable and regular motion, red ($10^{-3}$) to stochastic motion. Also we calculated the preservation of integrals of motion along the trajectories and presented the results in a similar way: blue colour corresponds to the jitter of $10^{-5}$, red $10^{-3}$. As one can see, there is good agreement between the two approaches but the FMA technique appears to be more sensitive.

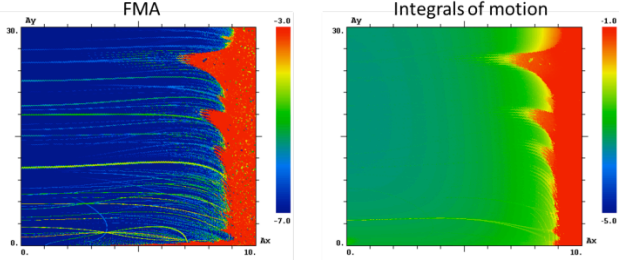

Figure 2: Simulations for IOTA with 4 nonlinear magnets, 20 magnet slices. Two plots in the plane of normalized betatron amplitudes ($10A_x \times 30A_y$) present the FMA (left) and preservation of the integrals of motion (right).

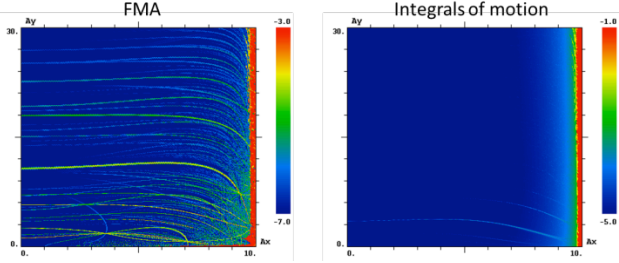

Figure 3: Simulations for IOTA with 4 nonlinear magnets, 200 slices per magnet.

The results of a similar simulation for nonlinear magnets split by 200 slices are shown in Fig. 3. In this case the integrals of motion are preserved with much higher accuracy, and the border of stable region is expanded horizontally up to the location of the magnet pole (and the singularity of potential). Nevertheless, at small vertical amplitudes there is no change in the FMA plot showing stochastic behaviour. A detailed inspection showed that in the vicinity of resonances, betatron tunes calculated within a window of 1024 turns can be modulated with some low frequency when the window shifts along the trajectory. The FMA algorithm interprets this behaviour by as tune jitter, while actually it is regular motion. The examples of possible tune dependence on the window shift are shown in Fig. 4, where the case (b) corresponds to one of the points in the bottom-left corner of Fig. 2.

Thus we conclude that the resonances seen in Fig. 2 and 3 within the rectangle of $5A_x \times 15A_y$ (our area of interest) are not *real* in the sense that the motion remains to be regular and integrable. Another conclusion is that splitting the nonlinear magnet by 20 slices is quite sufficient for our purpose. Later on we rely mainly on FMA since this method works without any knowledge about the exact analytic expression for integrals of motion, which sometimes need to be modified in a way we do not know a priori. Since the case of 200 slices corresponds very closely to the ideal integrable system, we use it as the base line for further studies and claim that all "new" resonances appearing in FMA plots due to various perturbations are real, like the case (a) in Fig. 4.

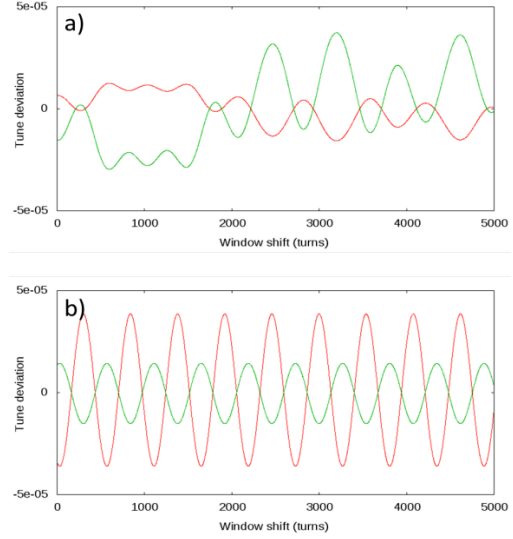

Figure 4: Betatron tunes deviation from the average vs. the window shift along the trajectory: real perturbation resulting in a tune jitter (a), and regular motion with tune modulation (b).

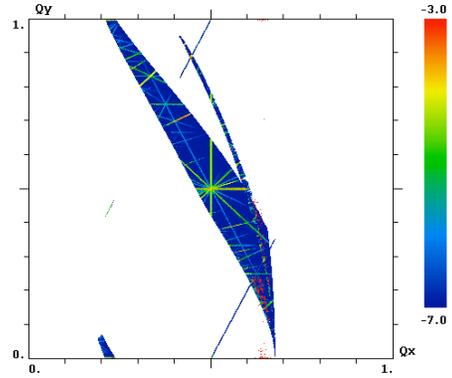

Figure 5: Footprint for IOTA with 4 nonlinear magnets.

The betatron tune footprint for IOTA with 4 nonlinear magnets is shown in Fig. 5. Here we considered only particles with amplitudes within the $5A_x \times 15A_y$ rectangle. When one of the amplitudes is small, it can be difficult to define the main frequency, which sometimes jumps to a nearest strong resonance. This effect results in the appearance of characteristic thin red lines in the footprint, which we consider as artefacts. The main body of the footprint corresponds to real betatron tunes, and two remarkable features can be seen in this simulation: a) a part of the footprint it crosses the vertical integer resonance; b) the vertical tune spread is larger than 1.

One of the possible options for the beginning of IOTA experimental program is the use of conventional octupole magnets in place of the special nonlinear magnets. The strength of octupoles must be equal to the corresponding term in the multipole expansion of the ideal nonlinear magnet, which results in the conservation of one of the integrals of motion. Simulation results for this case are shown in Fig. 6. The approximation of full nonlinear potential with one multipole harmonic apparently defeats integrability leading to the appearance of the dynamic aperture limitation: the white area in the space of normalized amplitudes (Fig. 6, right) corresponds to particles, which were lost within 2000 turns of tracking. However, for normalized amplitudes $A_{x,y} \leq 7$ the motion is stable, regular, and provides a large spread of betatron tunes, as is seen in Fig. 6 (left).

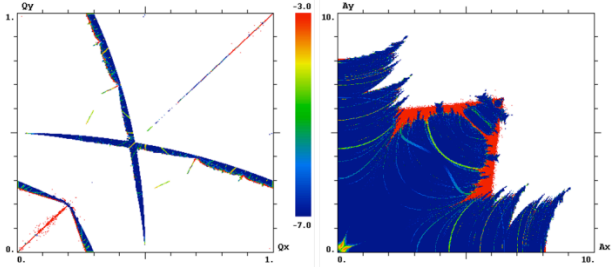

Figure 6: FMA plots for IOTA with nonlinear magnets replaced by octupoles: footprint (left) and normalized betatron amplitudes (right).

## MISALIGNMENTS AND OTHER IMPERFECTIONS

Misalignments of thin nonlinear magnets (slices) were simulated as random shifts in X and Y directions with r.m.s. of 10μ. Despite the appearance of many resonances and the expansion of stochastic layer close to the working area, the long-term tracking showed that no particles were lost after $10^6$ turns. However, with additional imperfections the situation becomes worse. In Fig. 7 the following effects were taken into account (all values are r.m.s): transverse misalignments 10μ, tilts 1 mrad, errors in magnet geometry (distance between the poles) $10^{-3}$, nonlinear magnet strengths $10^{-3}$, betatron phase advances in the linear lattice $10^{-3}$, beta-functions in the magnet section $10^{-2}$. As a result, about 6% of particles were lost after $10^6$ turns, which is tolerable considering the strength of nonlinearity and the magnitude of nonlinear tune shift. However, the simulation demonstrates that the error margins must be tightly controlled in the machine design.

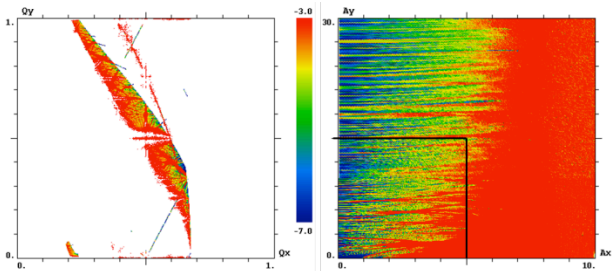

Figure 7: FMA plots for IOTA: misalignments, tilts, errors in geometry, magnets' strengths, betatron phase advances, beta-functions. The working area of $5A_x \times 15A_y$ is shown by black lines.

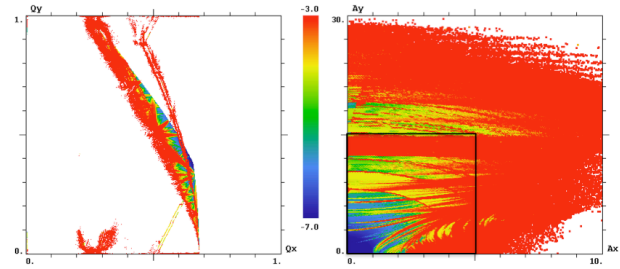

Figure 8: FMA plots for IOTA: full 6D simulation.

The above simulations were performed in the 4D approximation, disregarding a large class of imperfections disturbing integrability - the effects related to synchrotron motion. These originate from a non-zero dispersion in the nonlinear magnet, or from the non-isochronous particle oscillations. The former can be contained by the lattice design while the latter becomes especially pronounced in a lattice with small momentum compaction factor $\alpha$. In the ultimate IOTA configuration with 4 nonlinear magnets and dispersion fully suppressed, $\alpha$=0.015 and the momentum offset for a particle with the transverse amplitude of 25 mm becomes very large, approximately $5 \times 10^{-3}$, and the longitudinal dynamic aperture shrinks (particles leave the RF bucket). Because the effect of non-isochronous oscillations is quadratic in betatron amplitude, the natural cure for this issue is to decrease the latter by decreasing the magnet aperture. Fig. 8 shows the 6D FMA results for IOTA with the maximum betatron amplitude limited to 10 mm (magnet aperture 6.6 mm), and this simulation demonstrates that a tune spread close to 1 is achievable. Alternatively, one can increase momentum compaction at the cost of reducing the number of magnets from 4 to two or one (and lower total tune spread). That way the mechanical magnet design is not so challenging but still allows to attain the tune spread of 0.3 per cell.

## ACKNOWLEDGMENTS

The authors thank V. Kashikhin for discussions on the design of nonlinear magnets, J. Leibfritz and L. Nobrega for help with the ring design. We are grateful to V. Shiltsev for his continuous support of this activity.